\documentstyle[12pt]{article}
\input epsf

\if@twoside
 \else
 \fi

 \parskip \medskipamount

\begin{document}
\hfill{LA-UR-94-2863}

\vspace{7pt}
\begin{center}
{\large\sc{\bf Fermionic Operators from Bosonic Fields
in 3+1 Dimenions.}}

\baselineskip=12pt
\vspace{35pt}

A. Kovner$^ *$\\
\vspace{10pt}
Theory Division, T-8,
Los Alamos National Laboratory,
MS B-285\\
Los Alamos, NM 87545\\
\vspace{20pt}
B. Rosenstein$^ **$\\
\vspace{10pt}
Institute of Physics, Academia Sinica\\
Taipei, 11529\\
Taiwan, R.O.C.\\

\vspace{55pt}
\end{center}
\begin{abstract}
We present a construction of fermionic operators in 3+1 dimensions
in terms of bosonic fields in the framework of $QED_4$. The basic
bosonic variables are
the electric fields $E_i$ and their conjugate momenta $A_i$.
Our construction generalizes the analogous constuction of fermionic
operators in 2+1 dimensions. Loosely speaking, a fermionic operator
is represented as a product of an operator that creates a pointlike
charge and an operator that creates an infinitesimal t'Hooft loop
of half integer strength. We also show how the axial $U(1)$
transformations are realized in this construction.

\end{abstract}

\vspace{50pt}
\
\newline
*KOVNER@PION.LANL.GOV
\newline
**BARUCH@PHYS.SINICA.EDU.TW
\pagebreak

The problem of bosonization in dimension higher than two is a long
standing one. Many attempts have been made to generalize the 1+1
dimensional Mandelstam \cite{cm} construction to higher
dimensions \cite{earlybos}.
However, they all have an unattractive feature that the bosonized
theories turn out to be nonlocal \footnote{The term "bosonization"
is sometimes understood as
obtaining an effective low energy description of fermionic theories
in terms of bosonic fields. This in many cases is a straightforward
excersise \cite{burgess}. This is not what we mean by bosonization
in this paper.
Here we have in mind an exact mapping between the fermionic and
bosonic
Hilbert spaces, which should give an exact bosonic description of a
fermionic theory at all energy scales.}
Recently, using certain dynamical assumptions, one of us  and
P. Kurzepa
succeeded to perform
bosonization of QED in 2+1 dimensions \cite{kk}. The resulting
bosonic theory was found to be local and strongly interacting.
As
expected, in 3+1 dimensions the corresponding construction is
significantly
more complicated.
In this letter we perform a first step of the analogous program
in 3+1 dimension: the construction of fermionic operators in terms
of bosonic fields.

We consider quantum electrodynamics with one Dirac fermion in 3+1
dimensions. Our basic philosophy and motivation to consider this
particular theory
are basically same as in 2+1 dimensions.
They are described in  detail in ref.\cite{kk} and need not be
repeated here. Apart of obvious specific 3+1 dimensional
features, like  a nonabelian representation of the rotation group
and different discrete symmetries, note that now the
axial anomaly is present. This is different from 2+1 but reminiscent
of 1+1 QED. We elaborate later on its implementation.
We will work in
the Hamiltonian formalism, and find the temporal gauge ($A_0=0$)
to be most convenient to our purposes. The Hamiltonian is
\begin{equation}
H=\frac{1}{2}E^2+\frac{1}{2}B^2+\bar\psi\gamma^i(i\partial_i+eA_i)
\psi+m\bar\psi\psi
\label{ham}
\end{equation}
It is supplemented by Gauss's constraint,
\begin{equation}
\partial_iE_i=e\psi^\dagger\psi
\label{const}
\end{equation}
We take the same approach as in ref.\cite{kk}. We would like to
construct  gauge invariant fermionic fields in terms of bosonic
fields $E_i$ and their conjugate momenta, which would solve the
constraint eq. \ref{const}. In principle, as a next step we would
like to calculate fermionic bilinears, including the components of
the energy momentum tensor, in terms of these bosonic fields, and
thereby bosonise the theory. In this paper we will only perform the
first step of this program, namely construct the fermionic operators.

The fields $\psi_\alpha$ that appear in eq.\ref{ham} are not gauge
invariant. We therefore define
\begin{equation}
\psi^{CG}_\alpha(x)=\psi_\alpha(x)\Phi(x)
\label{cg}
\end{equation}
where,
\begin{equation}
\Phi(x)=e^{ie\int d^2y e_i(y - x)A_i(y)}
\label{fi}
\end{equation}
and
\begin{equation}
e_i(x)=-\frac{1}{4\pi}\frac{x_i}{|x|^3}
\end{equation}
is the electric field of a point like charge. The operator defined this
way is gauge invariant, and coincides with the original fermionic field
in the Coulomb gauge, in which the phase factor $\Phi(x)$ in
eq.\ref{fi} becomes unity.

Our ansatz for the bosonic form of the fermionic operator is inspired
by the results of ref.\cite{kk}. In 2+1 dimensions the fermionic
operators had the following basic structure\footnote{In this formula
we have not made explicit the
difference between the first and the second components of the Dirac
spinor. The details of the construction are given in ref.\cite{kk}.}:
\begin{equation}
\psi_{2+1}(x)=\lim_{|\epsilon|\rightarrow 0}\int d\hat\epsilon
\Gamma(\hat\epsilon)V^{1/2}(x+\epsilon)\Phi(x) V^{*1/2}(x-\epsilon)
\label{2+1}
\end{equation}
The different factors in this expression have a clear physical meaning.
The operator $\Phi$ is the exact $2+1$ dimensional analog of the
operator defined in eq.\ref{fi}. It creates the electric field of a
point electric charge at the point $x$. The role of this factor is to
ensure the correct transformation properties of $\psi$ under the
global electric charge transformation. The operator $V(x)$ ($V^*(x)$)
is the
operator that creates a magnetic vortex with the elementary flux
$2\pi/e$ ($-2\pi/e$). This is a local scalar field in the framework
of $QED_{2+1}$ \cite{kre}. Its appearance in eq.\ref{2+1} is natural,
since bosonization should be akin to duality transformation, and
$V(x)$ is in fact the dual field in the theory \cite{marino}. The
fact that it is the square root of $V$ that appears in the equation,
makes the operator $\psi$ double valued. This is a necessary property,
since otherwise $\psi(x)$ could not have the correct spinorial
rotational
properties. It is also in agreement with general arguments of
ref. \cite{finkelstein}. $V^*$ also necessarily appears in the
same expression, since $\psi$ should not carry magnetic flux,
and therefore the flux that is created by $V$ is cancelled by $V^*$.
This particular combination also insures anticommutativity of $\psi(x)$.
To make the product of local operators appearing in the definition of
$\psi$ well defined, one has to point split $V$
and $\Phi$. The integration over the direction of the point
splitting $\hat\epsilon$ with the appropriate phase
factor $\Gamma(\hat\epsilon)$ projects the composite operator
$\psi$ onto the spin $1/2$ representation of the rotation group.

To generalize this construction we should first understand what
is the generalization of the vortex operator $V$ to 3+1 dimensions.
The operator $V$ now should create an infinitesimally thin line of
magnetic flux. It will depend therefore on the curve. The explicit
construction has been given in refs.\cite{polchinski}, \cite{kr4}.
\begin{equation}
V(C)=\exp\left\{\frac{i}{e}\int
d^3y[a_i(y-x_c)E_i(y)+b(x_c-y)\partial_iE_i(y)]\right\}
\label{v}
\end{equation}
Here $a_i(x)$ is a vector potential of an infinitesimally
thin magnetic vortex along the curve $C$ :
$\epsilon_{ijk}\partial_j
a_k(x)=2\pi\tau_i(x,C)\delta^2_{perp}(x)$, where $\tau_i(x,C)$
is the unit tangent vector to the curve $C$ at point $x$,
and $\delta^2_{perp}(x)$ is the two dimensional delta function in
the plane normal to $\tau_i$. The function $b(x)$ satisfies
$\partial_i[b(x)]_{mod 2\pi}= a_i(x)$. Since $a_i(x)$ has a
nonvanishing curl, the function $b(x)$ must have a surface
of singularities ending at the curve $C$. For example, for a
straight line $C$ running along the $x_3$ axis one has
\begin{eqnarray}
a_i(x)&=&\epsilon_{ij}\frac{x_i}{x_1^2+x_2^2},\ \ i=1,2;\ \
a_3(x)=0 \nonumber \\
b(x)&=&\Theta(x)
\label{ab}
\end{eqnarray}
with $\Theta$ the polar angle
in the $x_1-x_2$ plane (Fig.1.).
\begin{figure}
\caption{ The function $\Theta(x)$.}
\label{f1}
\end{figure}
An alternative representation of $V$ is obtained by integrating by
parts the
$\partial_i E_i$ term in the exponential:
\begin{equation}
V(C)=\exp\{\frac{2\pi i}{e}\int_S d^2 s^i E_i+\frac{i}{e}
\oint_{|y|\rightarrow\infty} d^2s^i(b(x_c-y)E_i)\}
\end{equation}
where the integration is over the surface S of discontinuities of
the function $b$, which has the curve $C$ as its boundary, and the
second term in the exponential is the integral over the surface at
spatial infinity.
Note, that even though the surface $S$ enters formally the
definition of $V$,
the operator actually does not depend on it, but depends rather
only on the boundary $C$. This is the consequence of the fact, that
changing the surface changes the phase by $2\pi/e\oint ds^iE_i$,
which is an integer multiple of $2\pi$ due to quantization of
electric charge \cite{polchinski}.

Let us now define a standard set of operators $V_\eta(x,\hat n)$,
which are associated with the set of curves in Fig. 2.
Those consist of half a circle with the center at the
point $x$ and radius
$\eta$ in the plane specified
by the unit normal vector $\hat n$, and two semiinfinite segments
of a straight line in the direction of the third axis.
\footnote{We pick the third direction arbitrarily, for definiteness.
As will become clear later, none of our calculations depend on it.}
\begin{figure}
\epsfysize=3.5in
\caption{ The standard set of curves $C$.}
\label{f2}
\end{figure}
Note, that $V_\eta^*(x,-\hat n)$ is obtained from $V_\eta(x,\hat n)$
by a three dimensional parity transformation.
Note, also that the product $V_\eta(x,\hat n) V_\eta^*(x, -\hat n)$,
is an operator that creates a unit strength closed t'Hooft loop, in
the plane normal to $\hat n$, centered at $x$ with radius $\eta$.

We are now going to generalize the $2+1$ dimensional construction of
the Fermi operators, by basically changing the product of a
vortex and antivortex into an infinitesimally small
t'Hooft loop of half integer
strength.
Consider therefore the following operators
\begin{equation}
\chi(x)_{\hat n}= k\Lambda^{3/2}e^{-\frac{i\pi}{2e}\int d^3y
\partial_iE_i}V_\eta^{1/2}(x,\hat n)\Phi(x)V_\eta^{*1/2}(x, -\hat n)
\label{chi}
\end{equation}
The first exponential factor here is the same as in $2+1$ dimensions
\cite{kk}. We have also
introduced the ultraviolet cutoff $\Lambda$ so that $\chi$ has
a dimension of the canonical fermi field.
The constant $k$ in eq. \ref{chi} is should be chosen so as to
normalize correctly the fermionic bilinears (see \cite{kk}). The
operator $\Phi(x)$ is defined in eq. \ref{fi} in terms of
an operator $A_i$.
This operator $A_i$ should be
understood as a variable conjugate to $E_i$
\begin{equation}
[E_i(x), A_j(y)]=i\delta^3(x-y)\delta_{ij}
\label{com}
\end{equation}
We also require, that $\epsilon_{ijk}\partial_j A_k=B_i$.
In these respects $A_i(x)$ is similar to the original vector potential.
One should remember, however, that since we are working on the gauge
invariant subspace of $QED_4$, $A_i$ is a gauge invariant operator.
In this sence, it can be thought of as a vector potential in a
particular gauge. Up to now, we did not need to specify this gauge
condition. We will return to this point later.

The operator $V^{1/2}$ ($V^{*1/2}$) obviously has different locality
properties than the operator $V$. Whereas $V_C$ has support only on the
curve $C$, and the dicontinuity surface of the function $b(x)$ is
unphysical, this is not quite true for $V^{1/2}$ ($V^{*1/2}$). The
latter operator is double valued. By moving the surface of
discontinuities from $S_1$ to $S_2$, we multiply $V^{1/2}$ by a
phase factor $\exp\{\frac{i\pi}{e}\oint_{S_1-S_2} ds^iE_i\}$,
which due to quantization of electric charge can be either 1 or
$-1$. Fixing the surface of discontinuities, is therefore
equivalent to specifying what branch of the square root we choose
in the definition of $V^{1/2}$.

Although we do not have to specify this choice, and can just
think about $V^{1/2}$ as a single valued operator defined on a
double cover, it is sometimes convenient to visualize it as having
a definite surface of discontinuities. We will always think about
this surface as being parallel to the $[1,3]$ plane. Note, that it
is still true that $V^{*1/2}(x, -\hat n)$  the and $V^{1/2}(x, \hat n)$
are related by a parity transformation, and therefore the surface
of discontinuities associated with the former operator, is also the
parity transform of the surface of discontinuities associated with
the latter. Therefore, the surface of discontinuitites associated
with the closed loop, $V_\eta^{1/2}(x,\hat n) V_\eta^{*1/2}(x, -\hat n)$
in the limit of small $\eta$ is just the whole plane $[1,3]$.
This doublevaluedness, or effective nonlocality in the definition
of $V^{1/2}$ is directly responsible for the anticommutativity of
the operators defined in eq. \ref{chi}.
Using the fact that for the function $\Theta(x)$ defined in
eq. \ref{ab} satisfies
$\Theta(x-y)-\Theta(y-x)=\pm \pi$ for any two points $x$ and $y$,
we find:
\begin{equation}
\chi_{\hat n}(x)\chi_{\hat n'}(y)_{|x-y|>>\eta}=\chi_{\hat n'}(y)
\chi_{\hat n}(x)\exp\{i(\Theta(x-y)-\Theta(y-x))\}=
-\chi_{\hat n'}(y)\chi_{\hat n}(x)
\end{equation}
The anticommutation relations between $\chi(x)$ and $\chi^\dagger(y)$
also follow immediatelly for $x\ne y$. In fact, the canonical
anticommutator at coincident
points is obtained, quite generally, only in the limit where the regulator
in the definition of fermionic operators is removed $\eta\rightarrow 0$.
This is also true in other dimensionalities (\cite{cm}, \cite{kk}).
In this limit the operator $\chi$ does not depend on the direction
$\hat n$ and one obtains
\begin{equation}
\chi(x)\chi^{\dagger}(x)+\chi^{\dagger}(x)\chi(x) =2k^2\Lambda^3
\end{equation}
which for an appropriate choice of $k$ becomes a $\delta$ -
function in the continuum limit.

Complete analogy with $2+1$ dimensions would require at this point
to project the operators $\chi$ onto appropriate representation
of the rotation group.
However, before doing that we have to address one more question, which
was not present in $2+1$ d. This is the action on the fermionic fields
of the axial $U_A(1)$ transformation.
To implement the correct action of $U_A(1)$, we need to know the
form of the axial charge in terms of bosonic variables. We infer
this from the axial anomaly equation of $QED_4$.
\begin{equation}
\partial_\mu J_5^\mu=-\frac{e^2}{16\pi^2}
\epsilon_{\mu\nu\lambda\sigma}F^{\mu\nu}F^{\lambda\sigma}
\end{equation}

We take therefore the bosonic representation of the axial
current as:
\begin{equation}
J_5^\mu=-\frac{e^2}{8\pi^2}\epsilon^{\mu\nu\lambda\sigma}A_{\nu}
F_{\lambda\sigma}
\end{equation}
and in particular
\begin{equation}
J_5^0=-\frac{e^2}{4\pi^2}A_iB_i
\label{axial}
\end{equation}
This is completetly analogous to $1+1$ dimensional case.
We want to stress, that eq. \ref{axial} should not be thought of
as a guess, but rather as the equation defining what we mean by
the operator $A_i$. Up to now, we have only required that $A_i$
is a canonical conjugate of the electric field $E_i$,
and that $\epsilon_{ijk}\partial_j A_k=B_i$. Eq. \ref{axial}
defines also the longitudinal part of $A_i$ in terms of the gauge
invariant quantity $J_5^0$, in a way consistent with the anomaly
equation. This equation therefore can be understood as the gauge
condition referred to earlier.

Let us see what are the transformation properties of the operators
$\chi$ under the axial transformation. The operator $\Phi(x)$,
clearly commutes with $J_5^0$, since it only involves the momentum
$A$. The relevant commutator is, therefore
\begin{equation}
e^{i\alpha Q^5}V_\eta^{1/2}(x,\hat n) V_\eta^{*1/2}(x, -\hat n)
e^{-i\alpha Q^5}=V_\eta^{1/2}(x,\hat n) V_\eta^{*1/2}(x, -\hat n)
e^{i\frac{e}{2\pi}\alpha\phi_{\hat n}(x)}
\end{equation}
Here $\phi_{\hat n}(x)$ is the magnetic flux through the surface
bounded by the circle of radius $\eta$ with the center in $x$
lying in the plane with normal $\hat n$. In other words it is
magnetic flux through the area bounded by the closed t'Hooft loop
created by the operator
$V_\eta^{1/2}(x,\hat n) V_\eta^{*1/2}(x, -\hat n)$.

We can now define eigenoperators of the axial charge with
eigenvalues $\pm 1$
\begin{eqnarray}
\chi_{\hat n}(x)^{+}&=&\int d\alpha e^{i\alpha Q^5}
\chi_{\hat n}(x)
e^{-i\alpha Q^5}e^{-i\alpha}
=\int d\alpha \chi_{\hat n}(x)e^{i\alpha(\frac{e}{2\pi}
\phi_{\hat n}(x)-1)}
\nonumber \\
\chi_{\hat n}(x)^{-}&=&\int d\alpha e^{i\alpha Q^5}
\chi_{\hat n}(x)
e^{-i\alpha Q^5}e^{i\alpha}=\int d\alpha
\chi_{\hat n}(x)e^{i\alpha(\frac{e}{2\pi}\phi_{\hat n}(x)+1)}
\label{++}
\end{eqnarray}

The integral over the transformation parameter $\alpha$ can be
now easily performed, and we find
\begin{eqnarray}
\chi_{\hat n}(x)^{+}&=&\chi_{\hat n}(x)
\delta(\phi_{\hat n}(x)-1) \nonumber \\
\chi_{\hat n}(x)^{-}&=&\chi_{\hat n}(x)
\delta(\phi_{\hat n}(x)+1)
\label{++f}
\end{eqnarray}

This result has a very natural interpretation. The operator
$\chi_{\hat n}(x)^{+}$ is nontrivial only on states which
contain a unit magnetic flux
linked with the t'Hooft loop in the definition of
$\chi_{\hat n}(x)^{+}$.
Note, that the extra factor does not change the anticommutation
relations of the operators. This is obvious, since this extra
piece measures magnetic field at the location of the operator
$\chi(x)$, whereas the operator $\chi(y)$ creates only a closed
loop of flux at its location. More precisely
\begin{equation}
e^{i\alpha\frac{e}{2\pi}\phi_{\hat n}(x)}
V^{1/2}_{\eta}(y,\hat n')V^{*1/2}_{\eta}(y,-\hat n')=
V^{1/2}_{\eta}(y,\hat n')V^{*1/2}_{\eta}(y,-\hat n')
e^{i\alpha\frac{e}{2\pi}\phi_{\hat n}(x)}e^{i\frac{\alpha}{2}
l(C_1,C_2)}
\end{equation}
where $l(C_1,C_2)$ is the linking number of two small circles
with radius $\eta$ centered at points $x$ and $y$ with normals
$\hat n$ and $\hat n'$ repsectively.

We now project these operators onto spin $1/2$ representations
of the rotation group. We use the chiral representation of
the Dirac $\gamma$ matrices, in which each chirality component
transforms under rotations independently.
The complete Dirac multiplet is then constructed as
follows:
\begin{eqnarray}
\psi_1(x)&=&\int d\hat n Y_{1/2}(\hat n)\chi_{\hat n}(x)^{+} \nonumber \\
\psi_2(x)&=&\int d\hat n Y_{-1/2}(\hat n )\chi_{\hat n}(x)^{+} \nonumber \\
\psi_3(x)&=&\int d\hat n Y_{1/2}(\hat n )\chi_{\hat n}(x)^{-} \nonumber \\
\psi_4(x)&=&\int d\hat n Y_{-1/2}(\hat n)\chi_{\hat n}(x)^{-}
\label{psi}
\end{eqnarray}
Here $Y_{\pm 1/2}(\hat n)$ are monopole harmonics
\begin{equation}
Y_{\pm 1/2}(\phi,\theta)=\mp e^{\pm\frac{\phi}{2}}(1\mp \cos\theta)^{1/2}
\end{equation}
By construction therefore $\psi_1,\psi_2$ constitute a rotational
doublet of chirality $+1$, and $\psi_3,\psi_4$ - a rotational
doublet of chirality $-1$.

Eq.\ref{psi} is our final result. To conclude, we have constructed
in the framework of QED$_4$ the four dimensional multiplet
of fermionic operators entirely in terms of bosonic fields $E_i$
and their conjugate momenta $A_i$. The fermionic operators have
canonical anticommutation relations in the limit of zero regulator,
carry unit electric charge, can be decompose into two
representations of the axial charge with axial charges
$\pm 1$ and have correct transformation properties under
rotations. Note also, that under the parity transformation
$\phi(\hat n)\rightarrow -\phi(\hat n)$, and
$\chi_{\hat n}\rightarrow \chi_{-\hat n}$. Under the parity
transformation therefore the rotational multiplets $\psi_1, \psi_2$
and $\psi_3, \psi_4$ are transformed into each other, which is
precisely the right parity transformation properties of Dirac
fermions in the so called chiral basis of $\gamma$
matrices \cite{iz}\footnote{We wish to make a remark on the parity
transformation properties of the operators. If we were to
define the action of parity on $V^{1/2}$ as $V^{1/2}(\hat n)
\rightarrow V^{*1/2}(-\hat n)$ it would follow that $\chi
\rightarrow e^{i\frac{\pi}{e}\int d^y \partial _iE_i(y)}\chi$
which would not give the correct parity transformation properties.
However,  $e^{i\frac{\pi}{e}\int d^y \partial _iE_i(y)}$ is an
operator with eigenvalues $\pm 1$. Since $\chi$
is a double valued operator, we are free to change our
definition of parity so that $\chi\rightarrow\chi$.
This does not change the parity transformation properties
of all single valued operators and is therefore perfectly
consistent with the standard parity transformations of $A_i$
and $E_i$.}.

The bosonic form of the fermionic operators has a very clear
intuitive interpretation. The operator creates the electric
field of a pointlike electric charge and the magnetic field of
an infinitesimally small half integer magnetic fluxon. The extra
factor required to represent the axial charge, vanishes unless the
operator acts on a state which has a unit magnetic fluxon with a
linking number $\pm 1$ relative to the fluxon created by the
fermionic field itself.
The next logical step is to calculate fermionic bilinears in
 terms of the bosonic fields. This calculation was rather
involved already in $2+1$ dimensions, and is even more
complicated in the present case. Work along these lines is
currently in progress \cite{inprog}.

{\bf Acknowledgements} We thank P. Kurzepa for participation in
different stages of this work. A.K. is grateful to members of the
Institute of Physics, Academia Sinica for their kind hospitality
during his visit in November 1993, when part of this work was done.
B.R. is supported by the National Science Council ROC grant
NSC-82-0208-M-001-116

\end{document}